# Hybrid Femtosecond Laser and Ion-Implantation Processing for Controlled, Deep, High-Efficiency Ablation in Fused Silica

**Mario Garcia-Lechuga**[1,*], **Yoann Levy**[2], **Irene Solana**[1], **Fátima Cabello**[1], **María Dolores Ynsa**[3,4], **Nadezhda M. Bulgakova**[5]

[1] Laser Processing Group, Instituto de Optica Daza de Valdes (IO), CSIC, 28006 Madrid, Spain.

[2] HiLASE Centre, FZU - Institute of Physics of the Czech Academy of Sciences, Za Radnicí 828, 25241 Dolní Břežany, Czech Republic

[3] Departamento de Física Aplicada, Universidad Autónoma de Madrid, 28049, Madrid, Spain

[4] Centro de Microanálisis de Materiales (CMAM), Universidad Autónoma de Madrid, 28049, Madrid, Spain

[5] FZU - Institute of Physics of the Czech Academy of Sciences, Na Slovance 1999/2, 182 00 Praha 8, Czech Republic

*mario.garcia.lechuga@csic.es

# Abstract

Femtosecond laser modification of fused silica enables precise surface tailoring for the fabrication of micro-optical components such as microlenses and diffractive elements. However, the process is governed by laser–matter interactions where the local fluence determines the processing depth, often limiting control over feature geometry and efficiency. Here, we present a hybrid approach combining localized Au implantation (1.8 MeV $Au^{2+}$ ions) into $SiO_2$ samples with femtosecond laser irradiation (250 fs), effectively tuning the laser–matter interaction and resulting morphology. At both 515 nm and 1030 nm irradiation wavelengths, single-shot femtosecond pulses produce cylindrical craters with sharp edges and flat-bottom profiles. Independently of the fluence, these craters exhibit a constant depth of 550 nm, corresponding to the region of maximum Au concentration. The effect manifests already at moderate fluence (~4 J/cm²) and yields high ablation efficiency, up to 15 $\mu m^3/\mu J$. The hybrid method also works effectively at lower implantation doses that preserve the excellent transmission of fused silica, offering a promising pathway for the high-quality fabrication of flat optical components such as binary phase masks, phase lenses, or fused-silica micromolds.

# Introduction

Fused silica is a key material in optics [1]. Its high purity, broad transparency from the ultraviolet to the short-wave infrared, excellent chemical and thermal stability, low nonlinear refractive index, and high laser-induced damage threshold (LIDT) make it ideal for fabricating lenses, phase masks, and as a substrate for coated elements such as mirrors or filters. Beyond conventional glass manufacturing and mechanical processing, laser techniques have proven to be highly effective for fused silica tailoring and functionalization. They enable smooth and precise polishing [2], the creation of micro-structures or micro-optical elements on the surface [3,4] or subsurface [5,6], and, more recently, its additive manufacturing through 3D printing [7,8].

Beyond the aforementioned technological interest, fused silica is also widely used as a reference material for laser ablation and laser damage studies. Historically, advances in laser technology have prompted repeated assessments of its LIDT and ablation behavior [9–11]. More recently, with the advent of modern ultrashort-pulse laser systems, LIDT studies have been conducted across spectral ranges from the UV to the MIR [12,13] and at new repetition rate regimes, such as GHz-burst operation [14,15].

Regardless of the laser parameters, the final crater shape mainly depends on how the laser energy couples to the material. Without delving into detailed fundamental mechanisms, it is important to note that due to the transparent nature of fused silica, absorption starts through nonlinear processes that generate a free electron plasma. As a result, the material modification strongly depends on the local laser fluence, which governs plasma formation both at the surface and throughout the subsurface light propagation region [16]. In the most common scenario of spatial Gaussian excitation, the resulting surface transformation in fused silica represents a quasi-Gaussian crater, with sharper edges for higher nonlinearities in the absorption process [13]. This behavior holds for moderate laser fluence ranges [17]. However, at laser fluences well above the fluence threshold for modification, a flatter crater is formed [18,19] due to plasma effects that shield deeper regions from further excitation [20]. So, in summary, the crater characteristics (crater shape, maximum crater depth, and ablation efficiency) are determined by the underlying laser–matter interaction mechanisms, whose inherently complex and dynamic nature makes their prediction a theoretical challenge [21].

Nonetheless, the situation changes when processing fused silica thin films on silicon or other highly reflective and absorptive materials. For films around 100 nm thick, once a certain local fluence threshold is exceeded, complete thin-film removal occurs regardless of the fluence distribution [22]. This effect is attributed to an energy coupling on the silicon substrate, leading to a mechanical spallation of the layers [23]. For thicker dielectric layers, internal intensity maxima form due to interference between the incident and substrate-reflected beams, resulting in quantized ablation at depths corresponding to these maxima [24].

In this work, we propose a hybrid femtosecond laser processing approach for fused silica via prior localized Au ion implantation, enabling a fs-laser–matter interaction behavior similar to that of a thin film, but in a bulk material. While Au implantations have been commonly used in previous studies to investigate color control effects [25,26] rather than ablation processes, here we explore their impact on tailoring the surface modification $SiO_2$ samples. The formation of craters with sharp edges and a consistently flat bottom at a depth of 550 nm is demonstrated and discussed, after exploring different Au concentrations and excitation wavelengths. The obtained results, including ablation efficiencies, are systematically compared to the transformations observed in pristine fused silica.

**Materials and Methods**

**Sample fabrication and characterization**

We use fused silica wafers (JGS2 wafer from MicroChemicals GmbH) of 10 x 10 mm$^2$ and 500 µm thickness. Ion implantation is carried out using 1.8 MeV Au$^{+2}$ ions in a 5 MV linear tandem accelerator along the standard multipurpose beamline at CMAM [27]. The implantation is performed under static conditions over an approximately square area of $6.4 \times 6.4$ mm$^2$. The ion flux is monitored by measuring the beam current with a Faraday cup, and two samples are implanted at ion-irradiation fluences of $10^{15}$ and $10^{16}$ ions/cm$^2$ by adjusting the implantation time. To induce the nucleation of Au ions into gold nanoparticles (NPs) and to anneal the radiation-induced defects in the silica matrix, we perform thermal treatment following the procedure described in Ref. [26] for soda-lime glass. The samples are annealed in a furnace (Mestra, HP-25) with a heating rate of 10 ºC·min$^{-1}$ until reaching 800 ºC, where the temperature was held for 5 hours.

To characterize the Au-distribution in the sample, Rutherford Backscattering Spectrometry (RBS) measurements (2 MeV He$^+$ ions at normal incidence) are performed after annealing. Figure 1(a) shows the RBS spectrum for the sample implanted with a fluence of $10^{16}$ ions/cm$^2$. The RBS data indicate that Au is localized within a specific region inside the bulk material, which is composed of both Si and O. Using the software SINMRA 7 [28], these experimental data are fit by introducing 21 layers of 30 nm with different concentrations inside SiO$_2$. The best fit corresponds to the Au-concentration distribution shown in Figure 1(b). The maximum concentration reaches $3.3 \cdot 10^{20}$ at/cm$^3$ at 595 nm below the surface. The Au-distribution is fit with a Gaussian distribution of 260 nm full-width at half maximum. Similar measurements and analyses are carried out for the sample implanted with a fluence of $10^{15}$ ions/cm$^2$, obtaining the same Au-distribution but with a concentration one order of magnitude lower.

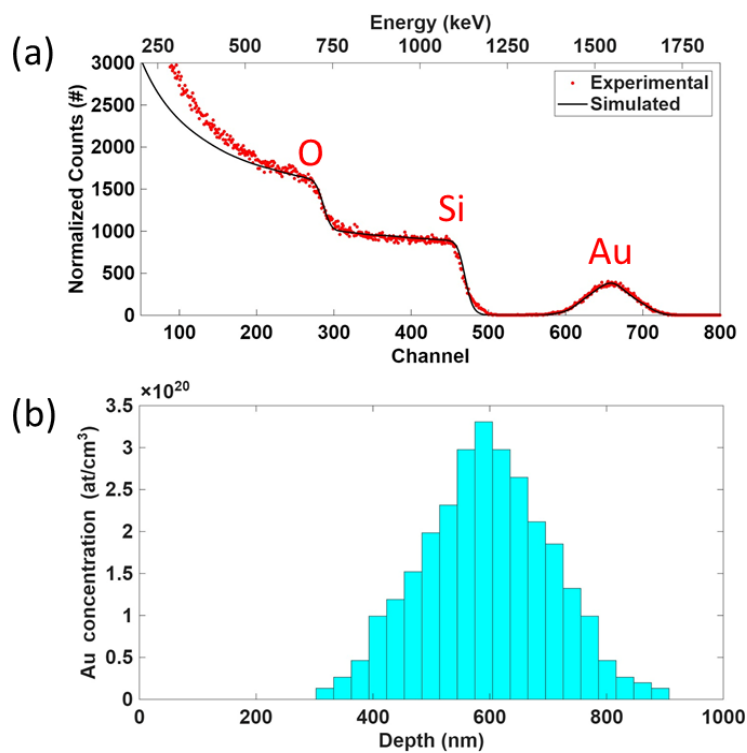

**Figure 1.** (a) (points) RBS spectrum of the Au-implanted fused silica under a dose of $1·10^{16}$ ions/cm$^2$. (line) Fit accounting for the volumetric distribution of Si, O, and Au. (b) In-depth Au-distribution as extracted from RBS spectrum.

Optical characterization of the samples is carried out with a spectrophotometer (Varian, Cary 5000 UV-Vis-NIR). Reflectivity and transmission spectra are measured for the pristine and the two implanted fused silica samples, as shown in Figure 2(a,b). With these measurements, the sample absorption is calculated and presented in Figure 2(c). On one hand, the higher-implanted sample displays the plasmon resonance of Au NPs close to λ=520 nm. On the other hand, the lower-implanted sample shows very low absorption (< 2%) and no visible indications of a plasmonic response from Au NPs.

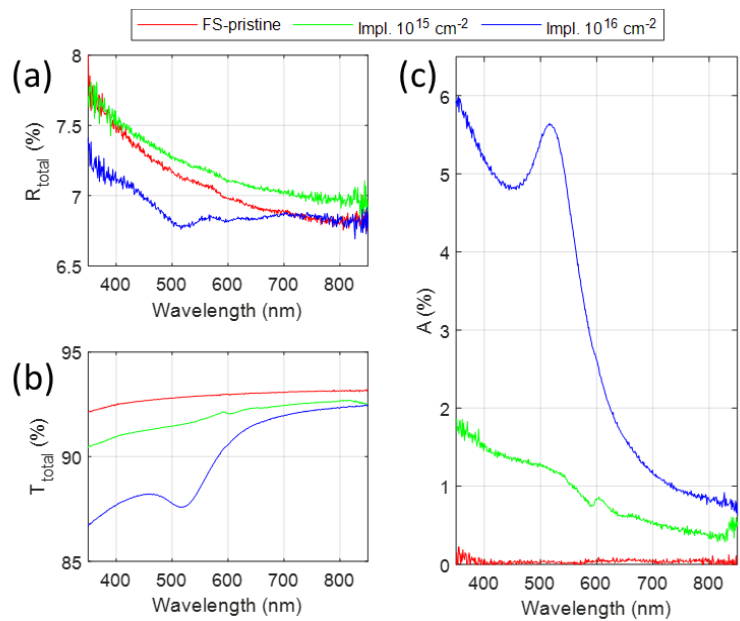

**Figure 2.** Optical properties of pristine and Au-implanted ($1\cdot10^{15}$ and $1\cdot10^{16}$ ions/cm$^2$) fused silica samples. (a) Total reflectivity measurements. (b) Transmission measurements. (c) Calculated absorption from reflectivity and transmission measurements.

**Femtosecond laser irradiation and crater characterization**

The femtosecond laser (PHAROS, Light Conversion) is coupled to a harmonic generation module (HIRO, Light Conversion), providing two separate outputs: a fundamental wavelength beam (1030 nm) emitting pulses at 260 fs pulse duration (FWHM) and the second harmonic beam (SH, 515 nm, 250 fs FWHM). Focusing conditions are adapted to have a similar spot radius ($w_0$) at the sample surface, determined using the Liu method [29] applied to laser-induced amorphous features produced in Si samples, as described in [30]. This yields $w_0 = 15.6$ µm for $\lambda = 1030$ nm and $w_0 = 13.3$ µm for $\lambda = 515$ nm.

Single-shot laser irradiations are done at different pulse energies ($E_p$). For $\lambda = 1030$ nm from $E_p$ was varied from 6.9 to 83.5 µJ, covering a range of peak laser fluences ($F_0$) of $F_0 = [1.8, 21.8]$ J/cm$^2$. For $\lambda = 515$ nm, the ranges are $E_p = [2.8, 58.3]$ µJ, and $F_0 = [1.0, 21.8]$ J/cm$^2$. All craters are characterized by optical methods, namely optical microscopy (Nikon Eclipse Ti, at 460 nm) and interference microscopy (Sensofar Plµ 2300, at 460 nm, Mirau-type 50x NA objective). The latter provides access to the crater morphology and volume. For a more accurate characterization of the crater topography, atomic force microscopy (AFM, Park Systems XE7) is used for a selection of craters. Radial analysis is performed on these images, allowing the association of each radial position ($r$) with a corresponding local fluence ($F(r)$) calculated as:

$$F(r) = F_0 \cdot e^{-2r^2/w_0^2} \qquad (1)$$

It should be noted that, from this analysis, the measured crater radius ($r_{crater}$) can be directly related to the ablation fluence threshold ($F_{th}$) through:

$$F_{th} = F_0 \cdot e^{-2r_{crater}^2/w_0^2} \qquad (2)$$

## Results and discussions

### Formation of cylindrical flat-bottomed craters

Figure 3 shows the modifications produced on fused silica samples, pristine and Au-implanted, upon irradiation by a single fs-pulse at 515 nm wavelength and $F_0 = 7.2$ J/cm². Optical microscopy images (Figure 3 (a-c)) already reveal a clear distinction between the pristine sample and the implanted ones. In the pristine sample, a reflectivity contrasting ring pattern is seen, associated with the formation of a heat-affected layer within the modified area [31,32]. In the implanted samples, sharp crater edges are observed.

AFM images shown in Figure 3 (d-f) confirm this difference. As analyzed in Figure 3(g), the crater profile in the pristine sample has a smooth crater depth transition toward the center of the laser-irradiated area, reaching approximately 200 nm depth. In contrast, both craters produced on the implanted samples exhibit a sharp topography change at the crater borders, followed by a nearly constant depth of around 600 nm. This cylindrical shape is atypical for a bulk material and represents a novelty of this work. It more closely resembles the peel-off behavior observed in thin films deposited on a substrate [22], exhibiting a flat-bottom crater profile and signs of partial layer detachment at the edges, visible for example along the left border in Figures 3(c) and 3(f).

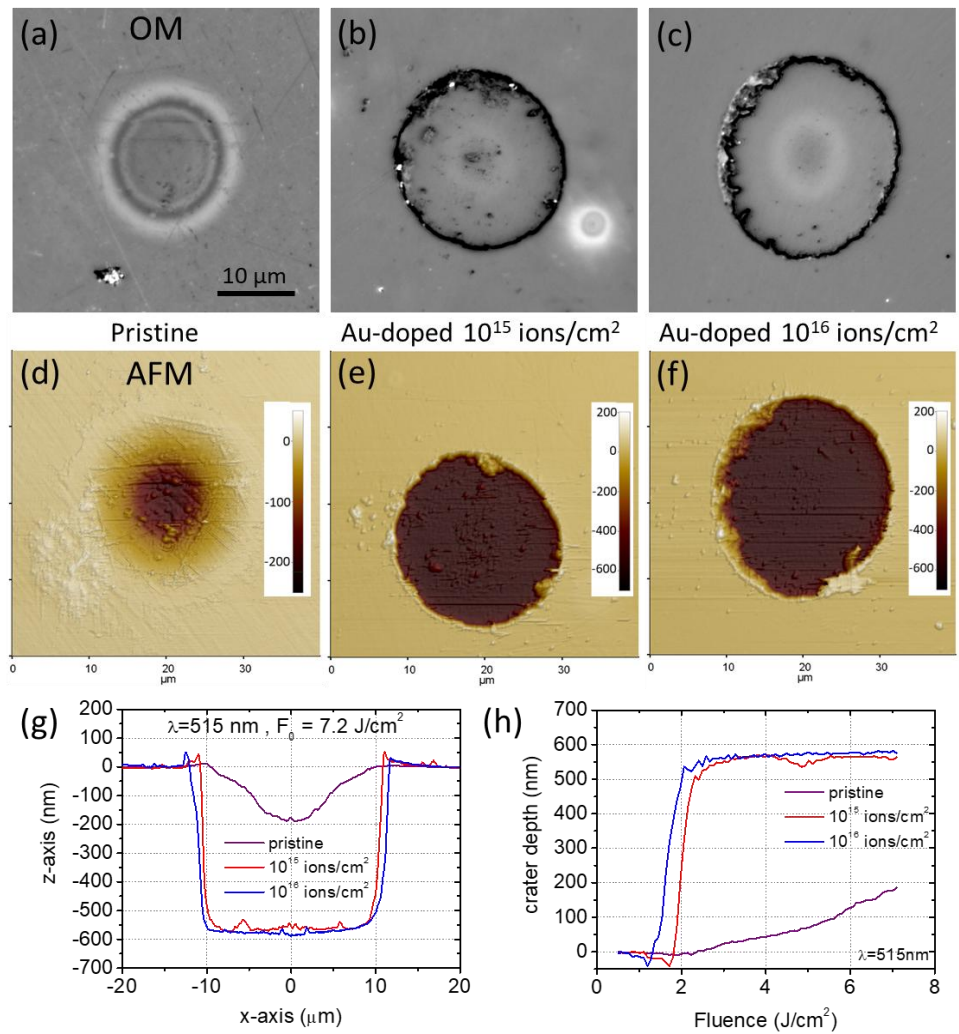

**Figure 3.** Laser-induced modifications in pristine and Au-implanted fused silica samples under a single-pulse laser irradiation (515 nm wavelength, 250 fs pulse duration, and 7.2 J/cm$^2$ peak laser pulse fluence). (a-c) Optical microscopy (OM) images recorded under a monochromatic illumination of 460 nm. (d-f) Corresponding AFM images. (g) Crater profiles extracted from the AFM images. (h) Crater depth dependence on the local fluence, as extracted from (g) after transforming the x-axis to local laser fluence (eq. 1).

Figure 3(h), where the x-axis is converted into local laser fluence values (Eq. 1), highlights other important aspects. It can be seen that for laser fluences close to 2.5

J/cm$^2$, which can be considered as the onset of ablation for the pristine sample, the implanted sample has already reached its maximum crater depth. Additionally, from these data, it follows that the lowest ablation fluence threshold is obtained for the highly Au-implanted sample. This result is consistent with the initial linear absorption at 515 nm, as shown in Figure 2, which reaches around 6%, while for the other Au-implanted samples, it is much lower (1%) or even zero for the pristine.

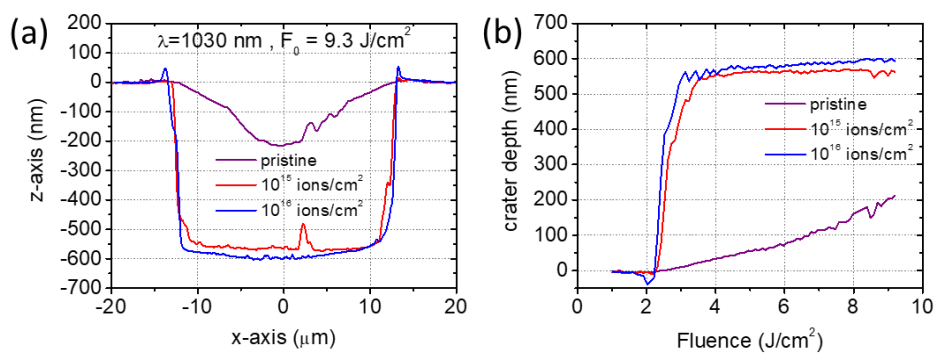

**Figure 4.** (a) Crater profiles extracted from the AFM images of modifications produced in pristine and Au-implanted fused silica samples under a single-pulse laser pulse (1030 nm wavelength, 260 fs pulse duration, and 9.3 J/cm$^2$ peak pulse fluence). (b) Crater depth dependence on the local fluence, as extracted from (g) after transforming the x-axis to local fluence (eq. 1).

Similar experiments were carried out using a 1030 nm laser wavelength, which corresponds to off-resonance radiation with respect to the Au plasmon. Figure 4(a) shows the crater profiles of the modifications produced for a peak laser fluence of $F_0$ = 9.3 J/cm$^2$. Similarly, as in Figure 3, the quasi-Gaussian shape for the pristine sample and cylindrical flat-bottomed craters in the implanted samples are observed. Furthermore, as shown in Figure 4(b), the maximum crater depth reached in the implanted samples approaches again 600 nm at moderate laser fluences (approximately 3.5 J/cm$^2$). In the literature, to reach those depths, Mirza et al. [19] needed to employ a peak laser fluence extremely above the fluence threshold, up to $F_0$ = 112 J/cm$^2$ (800 nm, 130 fs), and Jia et al. [33] used 5 pulses at values close to $F_0$ =

20 J/cm$^2$ (800 nm, 70 fs). Finally, and in contrast to the results at 515 nm, ablation fluence thresholds are similar for both implanted samples. This can be attributed to the fact that, since the irradiation is performed at an off-resonance wavelength, the linear absorption at 1030 nm does not differ significantly between the $10^{15}$ ions/cm$^2$ and $10^{16}$ ions/cm$^2$ samples.

In Table 1, a deeper analysis of the crater flatness characterized by AFM (Figures 3 and 4) is presented for the ion-implanted samples. Both the mean crater depth ($h_{mean}$) and the root-mean-square (rms) roughness were calculated considering only the crater area with depths below 400 nm, in order to exclude the crater edges from the analysis. All craters exhibit excellent flatness, with roughness values around 5% of the corresponding crater depth. Experimentally, we confirmed that the cylindrical, flat-bottomed crater shape is preserved across the full range of tested fluences. At the highest tested laser fluence, $F_0$ = 21.8 J/cm$^2$, the maximum crater depth in the implanted samples remains around 600 nm, independent of the laser wavelength. In contrast, under the same conditions, the pristine samples reach depths of about 320 nm for the 515 nm laser and 380 nm for the 1030 nm laser.

| Experimental conditions | | | AFM data analysis | |
|---|---|---|---|---|
| λ | $F_0$ | Implantation fluence | $h_{mean}$ ± rms | V |
| 515 nm | 7.2 J/cm$^2$ | $10^{15}$ ions/cm$^2$ | 549±29 nm | 195 μm$^3$ |
| | | $10^{16}$ ions/cm$^2$ | 553±35 nm | 235 μm$^3$ |
| 1030 nm | 9.3 J/cm$^2$ | $10^{15}$ ions/cm$^2$ | 546±31 nm | 245 μm$^3$ |
| | | $10^{16}$ ions/cm$^2$ | 567±38 nm | 290 μm$^3$ |

**Table 1.** Analysis of the mean crater depth ($h_{mean}$), root-mean-square roughness (rms), and crater volume (V) obtained from the AFM measurements shown in Figures 3 and 4.

After presenting the results for both laser irradiation wavelengths and examining two implantation levels, we now discuss the possible origin of the observed deep and controllable ablation process. As shown in Figure 1(b), the maximum gold concentration is found at a depth of approximately 595 nm, which closely matches the average crater depth observed in the implanted samples. Thus, the formation of craters appears to be directly linked to the region where the implanted gold atoms accumulate within the silica matrix. A plausible laser–matter excitation scenario is that absorption is most likely to be seeded in the region with the highest Au concentration, leading to a localized increase in carrier density through free-carrier absorption (inverse Bremsstrahlung) and impact ionization.

It should be noted that ion implantation leads to a change of refractive index of glass and its absorptivity due to several effects. (i) Ion bombardment causes damage and local densification of the glass lattice [34] with more atoms packed into the volume of maximum implantation that can enhance the effect of non-linear absorption. (ii) Even more important is the formation of color centers in fused silica glass upon ion implantation [35], which facilitate generation of free (seed) electrons upon laser irradiation [36] within the implantation volume. (iii) Although Au nanoparticles can be formed during the ion implantation stage that is strongly dependent on the ion energy and flux [37], the absorption is more determined by the implantation depth, where the major fraction of Au nanoparticles is located and, hence, more factors enhancing the local absorption are present. In this region, the incident laser energy is more efficiently absorbed. Au nanoparticles can further facilitate absorption in the surrounding $SiO_2$ matrix in their direct vicinity due to local field enhancement effects (particularly at 515 nm). All mentioned effects can contribute to the formation of a strong gradient in the free-electron density, which becomes increasingly preferential as more photons from the laser pulse arrive. This explanation accounts for why the craters do not exhibit either a shallower depth, since the energy is preferentially coupled beneath the surface, or a deeper one, as the absorption behind the implantation region drops.

Then, on longer time scales, the locally deposited energy is responsible for producing material heating and melting. We presume that, in our sample, this localized phase change generates a high-pressure regime beneath the surface, which may subsequently lead to spallation of the top layer. This scenario is inspired by the spallation observed in sub-hundred-nanometer dielectric thin films on silicon substrates [23], as well as by the phenomenon of quantized ablation reported in both thick dielectric films on semiconductors [24] and free-standing dielectric films [38]. In these cases, the localized phase changes occur within inner regions where constructive interference with the back-reflected surface enhances the local electromagnetic field, thereby defining preferentially excited zones.

Further time-resolved experiments will be conducted to investigate both the excitation and expansion dynamics, in order to determine their characteristic time scales and elucidate how the laser–matter interaction both resembles and differs from that observed in bulk dielectrics and thin-film ablation dynamics.

### Impact on ablation efficiency

As mentioned in the "Materials and Methods" section, laser irradiations were carried out over a wide range of energies and, consequently, peak laser fluences. The characterization of the full set of modifications is performed using interference microscopy. From those images, the ablated area of each crater is extracted by filtering pixels with depths higher than 10 nm. The depth criterion was used to avoid including topographical variations arising from the surface roughness itself. The ablated area data is used to determine the ablation fluence threshold ("the Liu method") for each experimental condition, finding for 515 nm wavelength values of 2.4 J/cm$^2$, 1.7 J/cm$^2$ and 1.4 J/cm$^2$ and for 1030 nm values of 3.1 J/cm$^2$, 2.5 J/cm$^2$, and 2.3 J/cm$^2$; respectively for pristine, implanted at $10^{15}$ ions/cm$^2$ and at $10^{16}$ ions/cm$^2$. Additionally, the ablated volume is determined by integrating the depth of each pixel of the crater. The characterization technique was validated by obtaining values that differed by approximately 10% from the ablated volumes measured by AFM in the same craters (Table 1).

In Figure 5(a,b), the volumes of all the produced craters are plotted. The main confirmation is the significant difference between the results for the pristine samples and the implanted ones. For example, in pristine samples, it is difficult to achieve ablated volumes of 100 µm³ even at high laser fluences, whereas in the implanted samples, this volume can be reached at laser fluences close to 4 J/cm² under all the conditions explored.

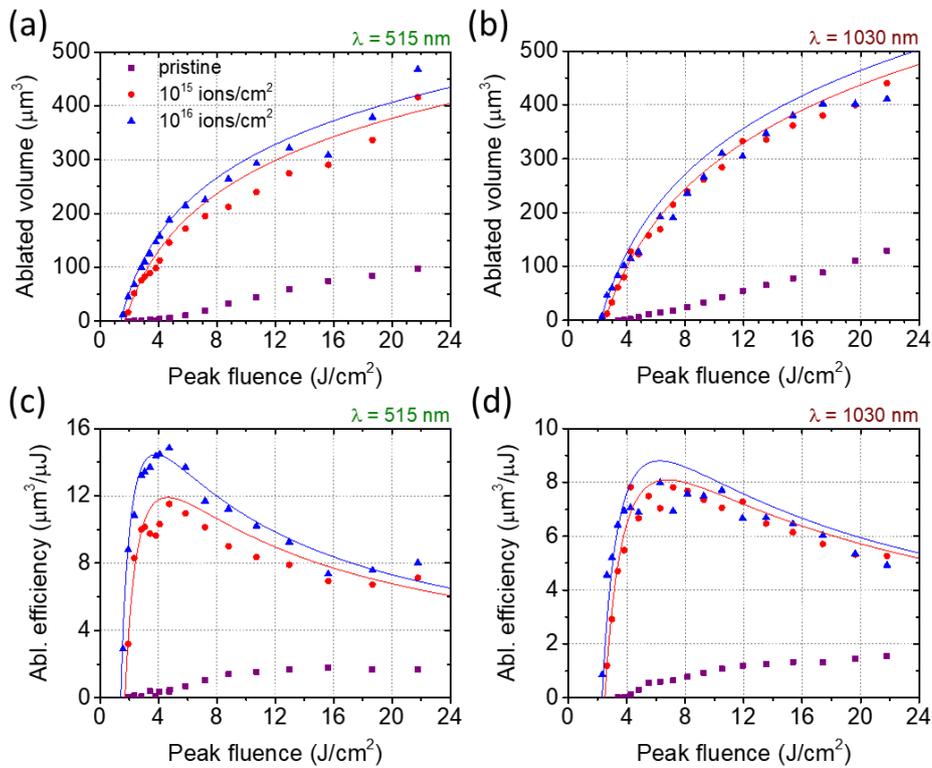

**Figure 5.** (a-b) Ablated volume of craters produced under single-shot irradiation at different peak laser fluences under 515 nm (a) and under 1030 nm (b) wavelength. Craters are characterized with an interference microscope. Solid lines represent the ablated volume of a perfect cylinder with a depth of h=550 nm (see eq. (3)). (c-d) Ablation efficiency calculated from the volume data plotted in (a-b) and divided by the corresponding pulse energy. Solid lines represent the ablation efficiency accounting for a perfect cylinder with a depth of h=550 nm (see eq. (4))

Together with the experimental data, Figure 5(a,b) shows the crater volume of a perfect cylinder ($V_{cyl}$) as a solid line. For this calculation, an average value of the constant depth observed in both Figures 3 and 4 ($h = 550$ nm) is used, and the crater radius ($r_{crater}$) is obtained from its relationship with the local laser fluence (eq. (2)):

$$V_{cyl} = \pi \cdot h \cdot r_{crater}^2 = \pi \cdot h \cdot \frac{w_0^2}{2} \cdot ln\left(\frac{F_0}{F_{th}}\right) \qquad (3)$$

The calculated $V_{cyl}$ serves as a useful reference to interpret the experimental trends. Despite a slight overestimation, likely due to the influence of the crater borders, which are not perfectly step-like and therefore deviating from the ideal cylindrical shape, the overall trend closely follows the experimental data. This agreement confirms that the data follow a logarithmic dependence on fluence. Furthermore, the increase in ablated volume for the more heavily implanted samples, especially at 515 nm, is well captured by the $V_{cyl}$ dependence on the fluence threshold $F_{th}$: lower threshold values lead to larger effective crater radii and, consequently, higher ablated volumes.

In Figure 5(c–d), the so-called ablation efficiency is shown, obtained by dividing the ablated volumes by the pulse energy. This parameter provides a normalized metric that can be used for comparison with other experimental data, independently of the spot size employed. Data show that the ablation efficiency for pristine material reaches a maximum value of 1.78 µm³/µJ for 515 nm ($F_0$ =15.6 J/cm²) and 1.54 µm³/µJ for 1030 nm ($F_0$ =21.8 J/cm²). This latter value agrees perfectly with the one obtained by Utéza et al. [18] also for infrared pulses (800 nm) and under a similar pulse duration (300 fs). Whereas for the implanted samples, under the most favorable conditions (515 nm, $10^{16}$ ions/cm²), the ablation efficiency increases up to 15 µm³/µJ. This value is one order of magnitude larger than other values reported in the literature for single-shot studies. For instance, Utéza et al. in their pulse-duration-dependent study (from 7 fs to 450 fs) reported a maximum ablation efficiency of 2.3 µm³/µJ, under pulses of 7 fs.

Furthermore, our results for all the conditions (equal or above 8 µm³/µJ) also appear to be comparable to studies involving multiple pulses, which generally enhance ablation efficiency by lowering the ablation fluence threshold through incubation effects or by promoting heat accumulation when irradiating at high repetition rates. For instance, the maximum ablation efficiency of 6.8 μm³/μJ reported by Yang et al. [39] under multipulse irradiation (1030 nm, 220 fs) at $f_{rep}$=50 kHz or the obtained value of 5.4 μm³/μJ from Metzner et al. [40] at $f_{rep}$=100 kHz (1030 nm, 190 fs). In principle, the only situations that surpass our results are those involving GHz-burst pulse interactions, such as the efficiencies approaching 50 μm³/μJ observed in [41] (GHz-burst of 10 pulses at $f_{rep}$=100 kHz, each pulse 230 fs) or the reported value of 26.8 μm³/μJ in [40] (GHz-burst of 8 pulses at $f_{rep}$=100 kHz, each pulse 190 fs).

Delving into the origin of these values and continuing the comparison with a perfect cylinder, the ablation efficiency for producing cylindrical craters ($\eta_{cyl}$) of constant depth *h* is:

$$\eta_{cyl} = \frac{V_{cyl}}{E} = \frac{h \cdot ln\left(\frac{F_0}{F_{th}}\right)}{F_0} \qquad (4)$$

which follows from Eq. (3) for $V_{cyl}$ together with $F_0 = 2 \cdot E/(\pi w_0^2)$. This relationship is plotted as continuous lines in Figure 4(c–d). As already shown for the volume, the experimental craters are well approximated by perfect cylinders; accordingly, the experimental ablation efficiencies are also well reproduced by the mathematical description $\eta_{cyl}$. Despite a slight overestimation, the maximum efficiency and the fluence at which they occur closely resemble the experimental data. Mathematically, these values can be obtained by differentiating Eq. (4) with respect to $F_0$ and setting the derivative to zero gives the maximum:

$$\eta_{cyl,max} = \frac{h}{F_{th} \cdot e} \qquad \text{at} \qquad F_0 = F_{th} \cdot e \quad (5)$$

with *e* being Euler's number. Therefore, as practical guidance: (1) Lower the $F_{th}$, the higher is the ablation efficiency, and (2) the most effective laser fluence is approximately three times the ablation fluence threshold. As a prove, in the experiment performed on the sample implanted with $10^{15}$ ions/cm² and irradiated with a laser wavelength of 515 nm, the peak ablation efficiency occurs at 4.7 J/cm$^2$, corresponding to about 2.8·$F_{th}$ under those conditions. The retrieved optimal fluence for the ion-implanted sample also highlights the enhanced effectiveness of the process compared with untreated dielectrics. For comparison, the Furmanski model predicts an optimal fluence of e$^2$·$F_{th}$ [42] and a recent refined modelling places the optimal fluence between 3 to 5·$F_{th}$ [43].

Also, as a perspective to be explored in future studies, this analysis opens the possibility of predicting in advance the crater shape, volume, and ablation efficiency for other Au-implantation depths. Since the implantation depth fundamentally determines the effective crater depth, $h$, and, together with the fluence threshold $F_{th}$, governs the entire ablation response, adjusting the implantation profile should make it possible to anticipate the ablation behavior.

## Conclusions

A hybrid process for the surface modification of fused silica has been demonstrated, combining prior deep Au-ion implantation with subsequent femtosecond laser irradiation (at both 515 nm and 1030 nm). By comparing the laser-induced transformations in pristine samples with those in Au-implanted fused silica, significant differences in both crater morphology and ablation efficiency were observed.

In pristine fused silica, the laser-induced craters exhibit a quasi-Gaussian profile, reflecting the influence of fluence-dependent laser-matter interactions. In contrast, in Au-implanted samples cylindrical craters are produced, showing sharp edges and a flat bottom with an average depth of 550 nm. Both the large depth and the crater shape are highly unusual for bulk dielectric materials and more characteristic of thin-film delamination effects. This behavior is attributed to preferential energy absorption at the depth corresponding to the maximum Au concentration.

As a result, the process achieves both a deep and predictable transformation, with highly efficient ablation, enabling substantial material removal at moderate laser fluence and significantly enhanced ablation efficiency. For all tested irradiation wavelengths and Au concentrations, the ablation efficiency reached values equal to or higher than 8 $\mu m^3/\mu J$, more than four times higher than that observed in pristine fused silica.

This hybrid method has the potential to enable the fabrication of ad-hoc designed binary phase optical elements (such as phase masks for beam shaping) or metasurfaces (e.g., photon sieves for strong focusing). Under the explored implantation conditions (dose of $10^{15}$ ions/cm$^2$), the process is expected to produce well-defined structures while having only a moderate impact on optical properties (< 2% transmission loss), highlighting its realistic applicability. Furthermore, based on the standard relationship for binary elements, $d = \lambda/[2(n - 1)]$, such structures could be designed for visible-light applications. By exploring implantation conditions closer to or deeper below the surface, this approach could potentially be extended to fabricate optical elements operating across other spectral ranges, demonstrating the versatility and future potential of the method for tailored photonic device fabrication.

**Acknowledgements**

This work is part of the projects ULS_PSB (PID2020–112770RB-C21 and PID2020–112770RB-C22) and HyperSpec (PID2023-148178OB-C22) funded by MICIU/AEI/10.13039/501100011033 and by ERDF/EU. It was partially funded by the CSIC through the call for internationalization i-LINK 2022 (project ILINK22056). The work of Y.L. and N.M.B. has been funded by a grant from the Programme Johannes Amos Comenius under the Ministry of Education, Youth and Sports of the Czech Republic SENDISO project No. CZ.02.01.01/00/22_008/0004596. Authors acknowledge the support from CMAM-UAM for the beam time (STD045/22, STD009/23, STD024/23 and STD025/23) and its technical staff support. I.S. acknowledges the support by the Spanish Ministry of Science, Innovation and Universities via a doctoral grant (FPU23/01300).

**Conflict of interest**

The authors declare no competing interests.

21 / 23